\documentclass[9pt,twocolumn]{article}
\usepackage{arxiv}

\usepackage[utf8]{inputenc} 
\usepackage[T1]{fontenc}    
\usepackage{hyperref}       
\usepackage{url}            
\usepackage{booktabs}       
\usepackage{amsfonts}       
\usepackage{nicefrac}       
\usepackage{microtype}      
\usepackage{lipsum}
\usepackage{multicol}
\usepackage{upgreek}
\usepackage{graphicx}
\usepackage{caption}

\title{Twisted Light and the mirage effect}

\author{
  Claire M. Cisowski$^{1}$ and 
 Ricardo R. B. Correia$^{1}$  \\ \\
 Instituto de Física, Universidade Federal do Rio Grande do Sul, Av. Bento Gonçalves 9500 Porto Alegre, RS, Brazil\\ \\
 Corresponding author: claire.cisowski@ufrgs.br
}

\begin{document}
\twocolumn[
  \begin{@twocolumnfalse}
\maketitle
    \maketitle
    \vspace{-0.3cm}
We study the transformation of an OAM-carrying beam propagating through a vertical gradient of refractive index present in a water-ethanol column. Proper liquid mixing and incidence conditions upon the solution cause the beam to undergo a mirror inversion along the vertical direction, forming an inferior mirage. This path reverses the handedness of the beam and unlike mirror inversion witnessed at sharp interfaces, mirror inversion occurs smoothly and topological charge inversion is accompanied by astigmatic changes. It is also possible to obtain, at the exit of the solution, a beam that has been partially inverted, thus presents spatially varying orbital fluxes.
\vspace{0.65cm}
  \end{@twocolumnfalse}]
  
\hspace{0.5cm} Since it was established that beams of light possessing an azimuthal phase dependence, also known as "twisted light", carry orbital angular momentum (OAM) \cite{PhysRevA.45.8185}, such beams have drawn considerable attention due to their remarkable properties and have found applications in a large variety of fields \cite{allen2003optical}. The OAM content of a beam can be manipulated by transforming the host beam \cite{e8e97f230fc848bc9c923a49b94e3f60}. Notably, mirror inversion causes handedness reversal, meaning that a beam carrying OAM in an amount of $\ell\hbar$ per photon, where $\ell$ is the topological charge of the optical vortex, will carry OAM in an amount of $-\ell\hbar$ per photon after mirror inversion. Handedness reversal is a key feature of various interferometric arrangements based on OAM-carrying beams \cite{Oliveira_2005,PhysRevLett.88.257901,PhysRevLett.96.113901} and is often carried out by Dove prisms, which rely on Total Internal Reflection \cite{Gonzalez:06}. Here, we study the transformation of an OAM-carrying beam propagating through a Vertical Gradient of Refractive Index (VGRIN) as it undergoes one mirror inversion along the vertical direction. In this case, the beam is deflected in the upward direction, forming an inferior mirage. Unlike in elements based on total internal reflection, mirror inversion occurs smoothly in a VGRIN, allowing to appreciate the dynamics of handedness reversal. Propagation of OAM-carrying beams in graded index media has mainly been investigated in cylindrical waveguides \cite{Hisatomi:05,Gregg:16}, where the gradient of refractive index maintains a form of azimuthal symmetry, it is not the case here. 

\begin{figure}[htbp]
\centering
\includegraphics[scale=1.0]{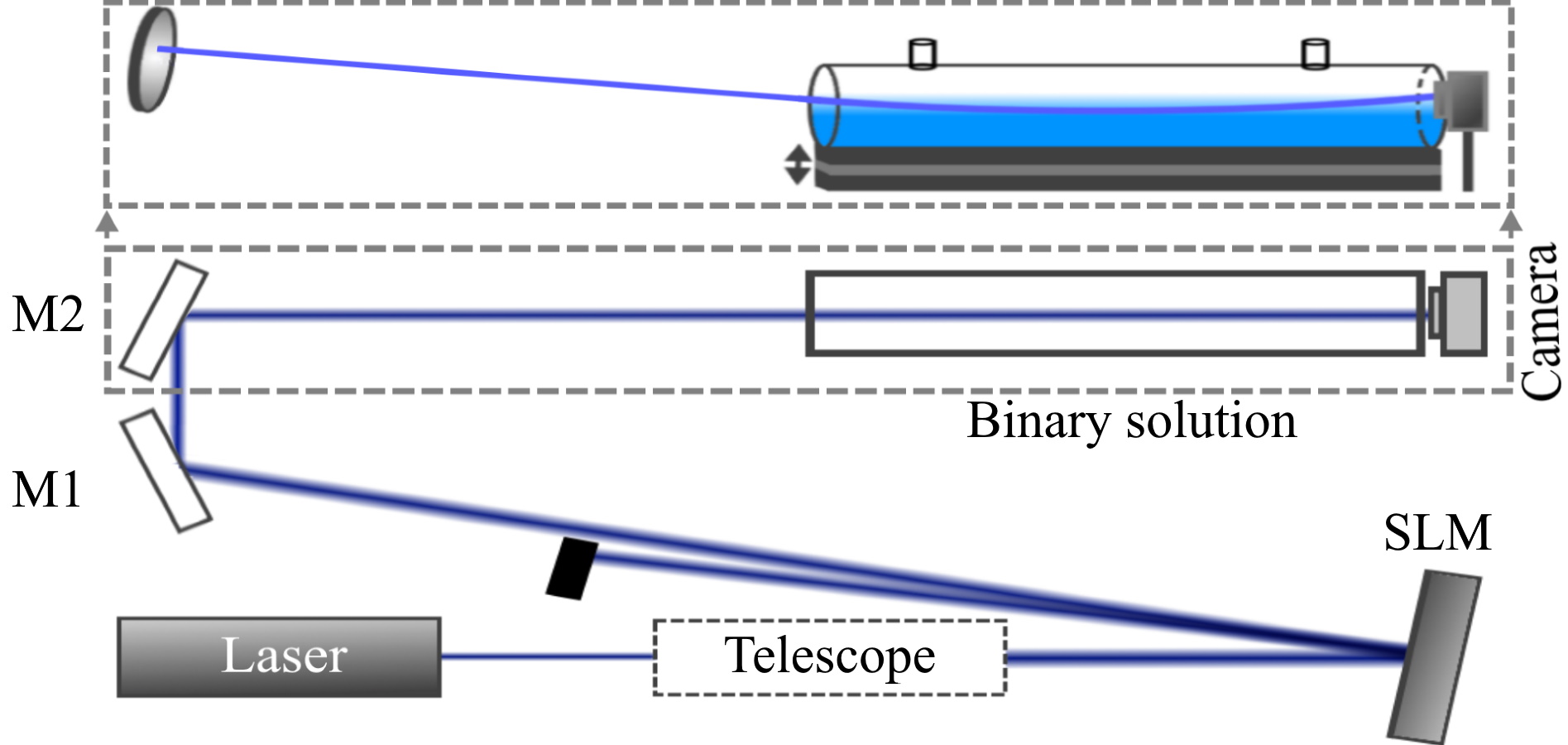}
\caption{Top view and side view (inset) representation of our experimental setup for studying the transformation of an OAM-carrying beam through a VGRIN.}
\label{fig1}
\end{figure}

 \hspace{0.5cm}We obtain a VGRIN by superposing liquids of different densities, a technique relatively risk-free, easy to implement, and which minimizes edge effects \cite{doi:10.1119/1.1986696}.
 We fill a cylindrical glass cell of inner diameter 26 mm and length 200 mm, enclosed with two quartz windows, with distilled water until the water level reaches 15 mm, then add ethanol (>99,9$\%$) until the cell is completely filled, while minimizing liquid mixing. Ethanol having a lower density and higher refractive index with respect to distilled water, a VGRIN establishes through the solution.
As illustrated on Fig.\ref{fig1}, we generate an OAM-carrying beam using a Fabry-Perot laser diode source of wavelength $\uplambda= 405$ nm, collimated through a telescope such that the waist diameter is $900\,\upmu\mathrm{m}$. A reflective phase Spatial Light Modulator (SLM), on which a fork diffraction hologram is displayed, imparts OAM to the first-order diffracted beam, which is spatially selected by blocking higher diffracted orders with an opaque mask. The selected beam carries OAM in an amount of $\ell\hbar$ per photon, where $\ell$ is the topological charge of the optical vortex, i.e., the number of $2\uppi$ phase windings around the phase singularity \cite{FREUND199999}. 
The incidence angle of the OAM-carrying beam upon the liquid solution is fixed to be $\uptheta_{\mathrm{in}}=1.5^{\circ}$ using a pair of plane mirrors (M1 and M2 on Fig.\ref{fig1}). The incidence height of the beam upon the solution is varied by placing the cell on a micrometric vertical stage. A CCD camera, of resolution 1296 x 964 and pixel size
3.75 $\upmu\mathrm{m}$, placed close to the cell exit plane, records the intensity distribution of the beam exiting the solution. 

\vspace{-0.25cm}

\begin{figure}[htbp]
\centering
\includegraphics[width=\linewidth]{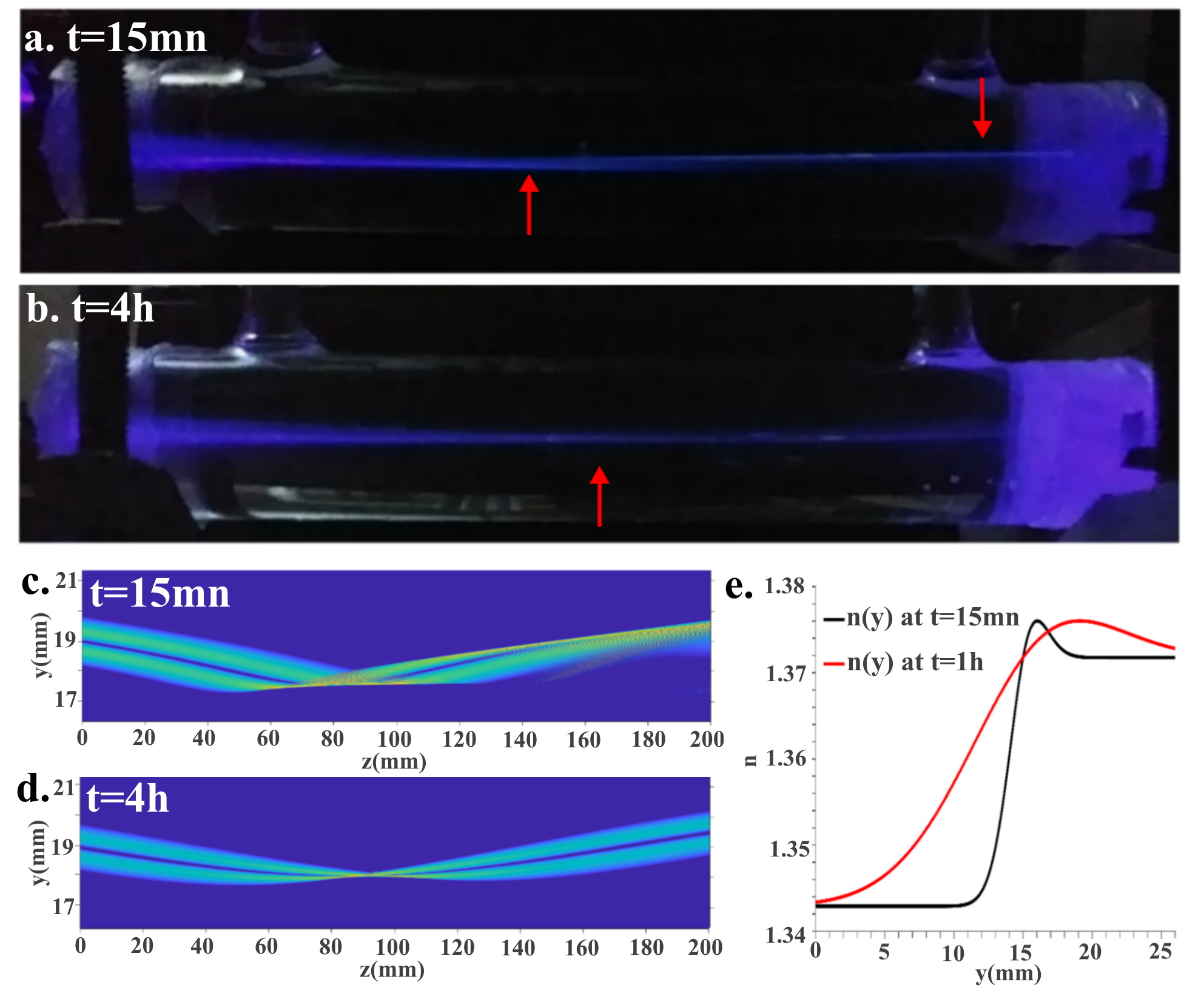}
\caption{Experimental (a.,b.) and numerical (c., d.) trajectories of a beam of topological charge $\ell=+1$ propagating through a binary solution of water and ethanol, for an incidence height $\mathrm{y}_{\mathrm{in}}=19$ mm and incidence angle $\uptheta_{in}=1.5^{\circ}$, 15 minutes (a., c.) and 4 hours (b., d.) after the cell has been filled. Deflection points are indicated by arrows on a. and b.. e. Refractive index of the solution as function of liquid height.   }
\vspace{-0.25cm}
\label{fig2}
\end{figure}

 \hspace{0.5cm}For an incidence angle of $\uptheta_{\mathrm{in}}=1.5^{\circ}$ and for an incidence height $\mathrm{h}_{\mathrm{in}}=19$ mm, i.e., 4 mm above the water-ethanol interface, deflections from the initial beam trajectory are visible to the naked eye, confirming the presence of a VGRIN within the solution. However, the beam trajectory changes over time: two inflection points are observed 15 minutes after the cell has been filled, whereas a single inflection point is observed after 4 hours have elapsed (see Fig. \ref{fig2} a.,b.). The VGRIN profile obtained in this binary solution thus evolves with time.
 \vfill\pagebreak

  \hspace{0.5cm}The temporal evolution of the VGRIN profile can be understood as follows. For $\uplambda=405$ nm and for a refractive index that does not vary along the horizontal direction, the refractive index of a binary mixture of water ($\mathrm{n}_{\mathrm{H_{2}O}}=1.3429$ \cite{Daimon:07}) and ethanol ($\mathrm{n}_{\mathrm{C_{2}H_{6}O}}=1.3725$ \cite{Kozma:05}) as a function of liquid height $\mathrm{y}$ can be expressed as \cite{ZHANG201414}:
\begin{equation}
\mathrm{n(y)}=\mathrm{n}_{\mathrm{H_{2}O}}+\mathrm{a}\cdot \upmu(\mathrm{y})+\mathrm{b}\cdot \upmu(\mathrm{y})^{2}+\mathrm{c}\cdot \upmu(\mathrm{y})^{3}
\label{eq1}
\end{equation}   
Where $\mathrm{a}=7.972 \times 10^{-2}$, $\mathrm{b}=-3.706\times 10^{-2}$, $\mathrm{c}=-1.381\times 10^{-2}$ and where \(\upmu(\mathrm{y})\) corresponds to the mass fraction of ethanol as function of liquid height, which can be written as \cite{Mandelis:84,Rashidnia:02}: 
\begin{equation}
\upmu{(\mathrm{y})}=(1+\mathrm{erf}(\mathrm{y}-\mathrm{y}_{0})/\mathrm{d})/2
\label{eq2}
\end{equation}
With $\mathrm{y}_{0}$ being the initial height of the water-ethanol interface and $\mathrm{d}$ being the diffusion length of this quasi-stationary distribution. The diffusion length evolves according to $\mathrm{d}^{2}=4\mathrm{Dt}$ where t is  measured in hours and where $\mathrm{D}=1.22\times 10^{-2}\mathrm{cm}^{2}\mathrm{s}^{-1}$  is the inter-diffusion constant of the two liquids \cite{doi:10.1021/je950222q}. As shown on Fig.\ref{fig2} e., diffusion decreases the steepness of the VGRIN over time. Also, a contraction of the water-ethanol mixture causes a peak in the refractive index profile, of higher value than the refractive index of the pure components alone \cite{LEE201362}, responsible for the secondary mirror inversion. Based on Eq. \ref{eq1} and Eq. \ref{eq2}, we plot, using BeamLab, the trajectory of a beam of topological charge $\ell=+1$ propagating though the solution, 15 minutes and 4 hours after the cell has been filled (Fig.\ref{fig2} c.,d.). Our numerical results are in agreement with our observations.
\vspace{0.1cm}

\begin{figure}[htbp]
\centering
\includegraphics[width=\linewidth]{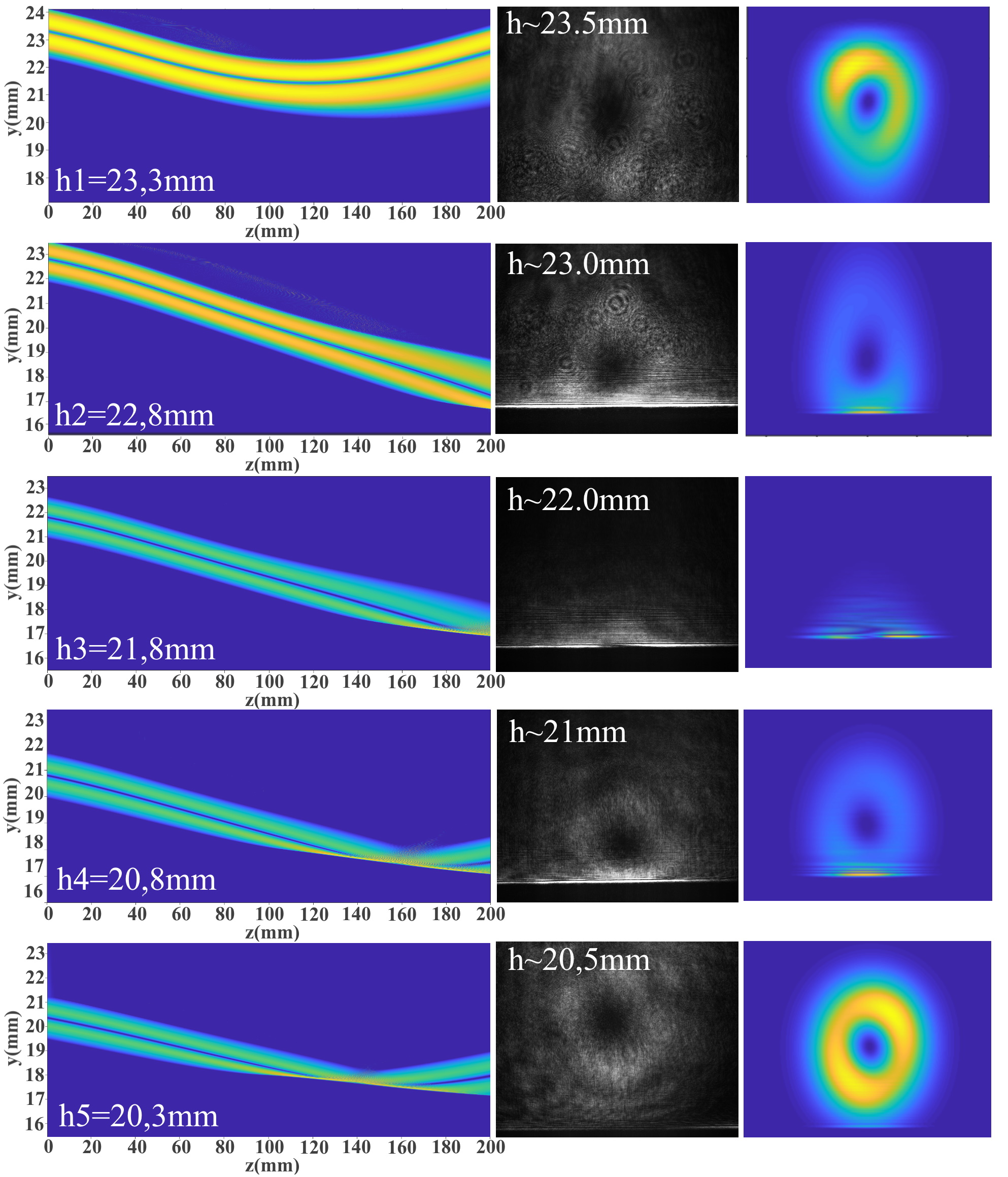}
\caption{ Numerical trajectory (first column) and experimental and numerical intensity distributions at the exit plane of the binary solution (second and third columns, respectively) of a beam of topological charge $\ell=+1$, for an incidence angle of $1.5^{\circ}$ and for various incidence heights, defined with respect to the center of gravity of the beam.}
\label{fig3}
\end{figure}

 \hspace{0.5cm}We are concerned with the transformation of an OAM-carrying beam experiencing a single mirror inversion, i.e. as an inferior mirage is formed. Thus, in what follows, we consider the VGRIN formed four hours after the cell has been filled, which is quasi-stationary ($<5$ $\%$ variations) over twenty minutes periods. Fig. \ref{fig3} presents the numerical and the experimental intensity distributions at the exit plane of the solution, for a beam of topological charge $\ell=+1$, entering the binary solution with an incidence angle $1.5^{\circ}$, as the incidence height is varied. The respective beam trajectories are also provided, to visualize how the beam transforms prior to reaching the exit plane of the solution. As the incidence height of the beam upon the solution decreases, the beam exiting the solution, initially non-inverted ($\mathrm{h}_{1}$), is partially inverted in the vertical direction ($\mathrm{h}_{2}$,$\mathrm{h}_{3}$,$\mathrm{h}_{4}$), until it is fully inverted ($\mathrm{h}_{5}$). For $\mathrm{h}_{3}$, only half of the incident beam is inverted, and both the inverted and non-inverted parts appear superposed at the exit plane of the cell. Both beam parts can be more readily distinguished upon free propagation, according to the vertical wavevector gradient introduced by the VGRIN. This gradient causes the beam to undergo astigmatic changes along the vertical direction, visible on Fig. \ref{fig3}, which are known to degrade the purity of the OAM mode \cite{2040-8986-15-1-014012}.
\vspace{0.1cm}

\vspace{-0.16cm}
\vspace{-0.25cm}

\begin{figure}[htbp]
\centering
\includegraphics[width=\linewidth]{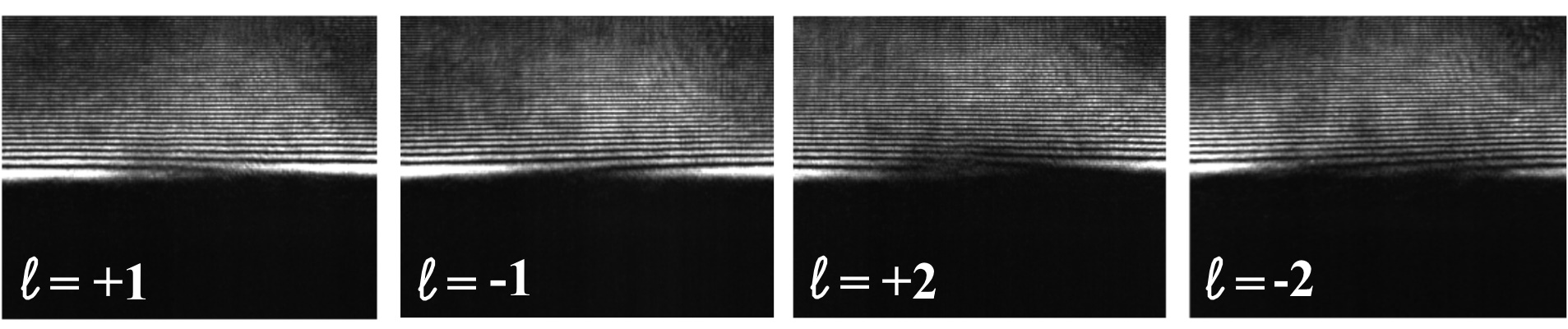}
\caption{Hermite polynomial-like modulation of the intensity distributions at the exit plane of the solution (experimental), for incident beams of various topological charges, as the optical vortex core undergoes mirror inversion ($\mathrm{h}_3$ on Fig.\ref{fig3}).}
\label{fig4}
\end{figure}

 \hspace{0.5cm}The astigmatic changes experienced by an OAM-carrying beam propagating through a VGRIN as it undergoes one mirror inversion are similar to the ones experienced by an OAM-carrying beam passing through the focal plane of a cylindrical lens, in a sense that mirror inversion occurs through continuous deformations of the beam along one direction  \cite{PhysRevLett.87.023902}. An Hermite polynomial-like modulation in the intensity distribution is observed where the optical vortex core of the incident beam undergoes mirror inversion ($\mathrm{h}_{3}$ on Fig.\ref{fig3}). As shown on Fig \ref{fig4}, this modulation is dependent on the topological charge of the incident beam, as it is usually the case in strongly astigmatic OAM-carrying beams \cite{VAITY20131154}.

\vspace{-0.25cm}

 \hspace{0.5cm} Mirror inversion can also be numerically studied by fixing the incidence height to obtain a trajectory like the one shown on Fig. \ref{fig2} d. and by varying the propagation distance of the observation plane. In this case the transformation of an OAM-carrying beam experiencing one mirror inversion are qualitatively the same, astigmatic changes occur along the vertical direction and a similar Hermite polynomial-like modulation is observed. Naturally, an OAM-carrying beam can also experience astigmatic changes that are non-related to mirror inversion as the beam follows U-shaped trajectories while propagating through the VGRIN, as showed on Fig. \ref{fig3} for the incidence height $\mathrm{h}_1$.
 \vspace{0.1cm}

\begin{figure}[htbp]
\centering
\includegraphics[width=\linewidth]{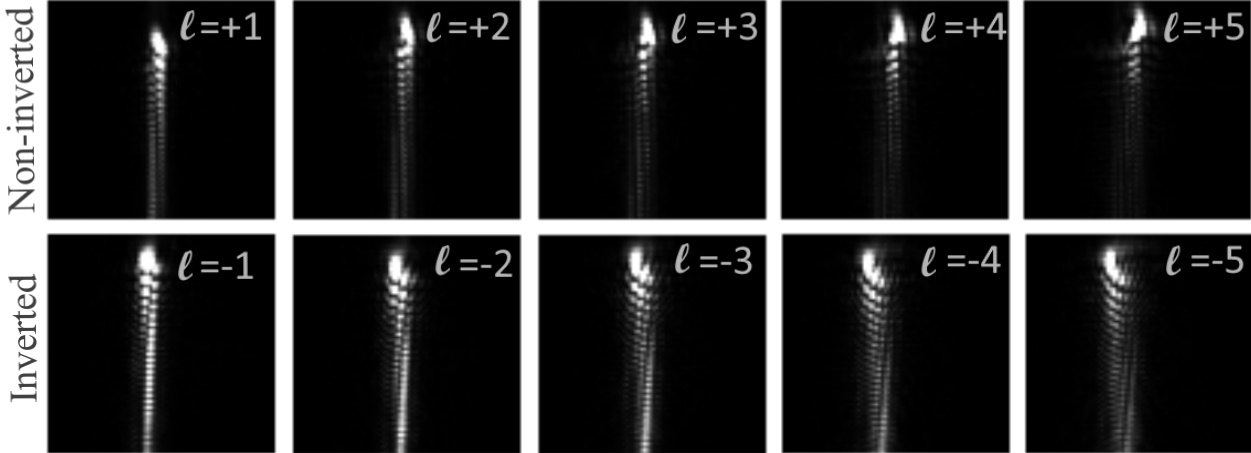}
\caption{ Intensity distributions, after propagating through a tilted convex lens, of OAM-carrying beams that have and have not experienced mirror inversion while propagating through the VGRIN. The number and orientation of the dark lines separating the bright lobes chains indicate the beam topological charge, written on each image.}
\label{fig5}
\end{figure}

 \hspace{0.5cm}We evidence handedness reversal upon mirror inversion by determining the average orbital angular momentum of a fully non-inverted and a fully inverted beam exiting the solution, for various incident topological charges. To do so, we record the intensity distribution of the freely propagating beam with a CCD camera, after it has passed through a tilted convex lens \cite{VAITY20131154}. As shown on Fig. \ref{fig5}, the difference of axial symmetry in the intensity distribution between a fully non-inverted and a fully inverted beam confirms that these beams are of opposite handedness.

 \hspace{0.5cm}For an observer at the exit plane of the solution, a partially inverted beam exiting the solution appears to carry spatially varying orbital fluxes, however, the exact OAM content of this beam is difficult to assess due to its inherent wave-vector gradient and deserves to be further investigated in the near-future.\vspace{0.1cm}
 
 \hspace{0.5cm}In summary, an OAM-carrying beam propagating through a VGRIN experiences continuous astigmatic changes along the vertical direction while undergoing mirror inversion, similar to the ones experienced by an OAM beam near the focal plane of an astigmatic lens. An Hermite-polynomial-like modulation in the intensity distribution is visible when the optical vortex core undergoes mirror inversion. This modulation is dependent on the topological charge of the incident beam. As expected, mirror inversion causes handedness reversal and at the exit plane of the solution and a partially inverted beam exiting the solution appears to have spatially varying orbital fluxes. The exact OAM content of this freely propagating beam can be difficult to assess due to the wavevector gradient in the vertical direction introduced by the VGRIN and deserves to be further investigated in the near future. Nonuniform liquid solutions capable of realising mirror inversion are more versatile than their bulk counterparts, we showed that various VGRIN profiles can be obtained within different period of time according to liquid mixing, also, optically active elements can easily be introduced, which is particularly interesting to study spin-orbit coupling between light and matter.

\section*{Funding}

This study was financed by the "Conselho Nacional de Desenvolvimento Científico e Tecnológico" and the "Coordenação de Aperfeiçoamento de Pessoal de Nível Superior - Brasil (CAPES) - Finance Code 001".

\bibliographystyle{unsrt}  
\bibliography{references}
\vfill

\end{document}